\documentstyle[aps,prl,epsf,twocolumn]{revtex}
\begin{document}
\def\bea{\begin{eqnarray}}
\def\eea{\end{eqnarray}}
\def\d{\delta}
\def\p{\partial} 
\def\#{\nonumber}
\def\r{\rho}
\def\la{\langle}
\def\ra{\rangle}
\def\e{\epsilon}
\def\break#1{\pagebreak \vspace*{#1}}
\draft
\title{Heat conduction in a one-dimensional gas of elastically
colliding particles of unequal masses}
\author{Abhishek Dhar $^{\dagger}$ }
\address{ Raman Research Institute,
Bangalore 560080, India. \\}
\date{\today}
\maketitle
\widetext
\begin{abstract}
We study the nonequlibrium state of heat conduction in a
one-dimensional system of hard point particles of unequal masses interacting
through elastic collisions. A BBGKY-type formulation is presented and
some exact results are obtained from it. Extensive numerical
simulations for the two-mass problem indicate that even for arbitrarily
small mass differences, a nontrivial steady state is obtained.
This state exhibits local thermal equilibrium and has a temperature
profile in accordance with the predictions of kinetic theory. The
temperature jumps typically seen in such studies are shown to be
finite-size effects. 
The thermal conductivity appears to have a very slow
divergence with system size, different from that seen in most other systems.
\end{abstract}


\narrowtext
{\it Introduction:}
The problem of finding a one-dimensional system of interacting
particles, evolving through Newtonian dynamics, in which  Fourier's law of
heat conduction holds, has attracted considerable interest in recent
years \cite{lebo}. One of the first studies in the field was the work
of Rieder et al \cite{ried} 
who obtained the exact nonequilibrium steady state (NESS) of a chain of
coupled harmonic oscillators. They found a constant temperature
profile in the system and a heat current that was
independent of system size. This result was not too surprising,
considering that the harmonic chain is integrable and
there is no scattering of phonons. Since
then, a large number of studies have looked at the 
effects of introducing impurities, nonlinearity and external
potentials [$3-8$].
The specific question asked in most of these studies is the
following. Defining the 
thermal conductivity $K$ of a system through the linear response formula 
$j=-K(T) \nabla T$,
where $j$ is the the current and $T$ the local temperature,   
is there a one-dimensional model which gives a finite $K$ ? 
Contrary to initial expectations, 
it was found that adding impurities and/or nonlinearity into
a system did not result in finite $K$.  The first
model in which a finite $K$ was found was the so-called ding-a-ling
model \cite{ding} in
which alternate particles on the line are bound to harmonic
springs and the other particles move freely between 
pairs of the bound particles. Numerical studies of several other models 
have also given finite conductivity \cite{bambi}. A common feature in these
models is the presence of external potentials which lead to momentum
nonconservation. 
Recently it has been proved rigorously \cite{prosen} that the
conductivity as given by the Green-Kubo formula 
always diverges in
one-dimensional momentum conserving systems with finite pressure.
Two recent studies \cite{giard} have reported finite conductivity in
momentum-conserving systems but both of these have vanishing pressure and so
there is no contradiction.

While the proof of anamalous conductivity in momentum 
\break{1.5in} 
conserving
systems has been a significant progress, 
there remain many issues that still need to be addressed. First of all, the
proof uses the Kubo formula 
and this is not fully satisfactory since the validity of the Kubo
formula, even in the limit of arbitrarily small temperature gradients,
has never been established rigorously. It may be noted that any
derivation of the Kubo formula for thermal conductivity essentially
makes the following assumptions: (i) the NESS is in
local thermal equilibrium (LTE) (ii) that Fourier's law is valid and (iii)
regression of equilibrium fluctuations occur in the same way as
nonequilibrium processes \cite{fors}. These assumptions are physically
motivated but may not be true in all cases.     
Secondly we note that while the focus of most of the work on heat
conduction has 
been addressed to obtaining Fourier's law, the more general problem
is one of understanding the nonequilibrium energy current carrying
state of a many-body system. 
For example one important question is the existence or otherwise of
LTE in the steady state. Thus one would
like to know if it is possible to define a slowly varying temperature
field which determines all other local properties in the system.
This point has not attracted  
much attention, even though it is quite crucial even for stating Fourier's
law and using  results of linear response theory.
Naively one might expect that if thermal currents in the system vanish in
the thermodynamic limit, then LTE should hold but this
has been shown not to be true always \cite{abhi}. 
Another interesting question is the determination of the temperature
profile itself. These questions are clearly of interest and can
be asked {\it independently of whether or not Fourier's law holds}. 
Finally we note that numerical studies of heat conduction are
problematic for several 
reasons. One needs long time numerical solution of nonlinear
differential equations which is time-consuming and not very
accurate. This, in addition to long equilibration times  typically 
occuring in such systems, restricts one to small system sizes. Also the
treatment of boundary related problems, such as that of temperature
jumps, is not straightforward. Hence
it has often been difficult to arrive at correct conclusions from
numerical studies and it is desirable to have more accurate studies. 

In this paper we study heat conduction in a system of hard elastic
particles of unequal masses moving on a one-dimensional line. The only
interaction between the particles is through elastic collisions. 
This model was first considered by Casati \cite{casa}
as a possible candidate for obtaining Fourier's law but the numerical
results were insufficient to draw any definite conclusions. 
More recently this model has been studied by Hatano \cite{hata} who
obtained a diverging conductivity. The equal-mass case with
dissipation was studied by Du et al \cite{du} who obtained a rather
surprising NESS which implied a breakdown of usual hydrodynamics. 

Here we present extensive and accurate numerical work on
this model and also an analytic formulation of the BBGKY-type.
Our aim has been to give a more detailed
characterization of the NESS than has been
previosly done.  
The present model is particularly suitable for this purpose for two
reasons: (i) simulations
of this model do not require a numerical solution of nonlinear
differential equations and it is possible to obtain very
accurate results, 
(ii) analytically the BBGKY hierarchy has a comparatively simpler
structure  and some exact statements can be made. 
The most interesting result obtained  is that the steady states for
the case of 
equal masses and the case with arbitrarily small mass differences 
are completely different. In the thermodynamic limit the latter case
exhibits LTE and the temperature profile approaches a 
form predicted by kinetic theory for a one-dimensional gas.This is
surprising since kinetic  theory predicts a {\it finite conductivity}
with a $T^{1/2}$ dependence on temperature.
On the other hand our model is momentum conserving
and the proof for diverging Kubo conductivity holds.  
In our finite size studies we find a slow
divergence of the conductivity ($\sim L^{\alpha}$ with $\alpha < 0.2$).
Our work also clarifies some of the
problems related to boundary effects. The jumps in temperature at the
boundaries are shown to be
finite-size effects which are studied systematically.

{\it Definition of model:}
We consider $N$ point particles numbered $i=1,...N$ moving in a
one-dimensional box extending from $0$ to $L$. The mass, position and
velocity of the $i$th particle are denoted by $m_i$, $x_i$ and
$u_i$. The only interaction between particles is through elastic
collisions. After a collision between particle $i$ and $(i+1)$, the
new velocities are obtained from momentum and energy conservation and
given by the linear equations:
\bea
u_i'=\frac{m_i-m_{i+1}}{m_i+m_{i+1}} u_i+ \frac{2 m_{i+1}}{m_i+m_{i+1}}
u_{i+1} \nonumber \\
u_{i+1}'= \frac{2 m_{i}}{m_i+m_{i+1}}
u_i-\frac{m_i-m_{i+1}}{m_i+m_{i+1}} u_{i+1}. 
\label{coll}
\eea 
Between collisions the particles travel with constant velocity. The
coupling to heat baths is implemented by using Maxwell boundary
conditions. Thus whenever a particle of mass $m$ collides with a wall at
temperature $T$, it is reflected back with a velocity chosen from the
distribution $P(u)= (m \mid u \mid /{T}) exp{(-m u^2/(2
T))} $. The temperatures at the two ends are taken to
be $T_1$ and $T_2$.

The
case when all particles have equal masses can be solved easily and
behaves similarly to the ordered harmonic springs case
\cite{ried}. The ``temperature'' 
profile is flat (with the value $\sqrt{T_1 T_2}$), current is
independent of system size and there is no 
LTE. The equal-mass problem is integrable and essentially reduces to a
single-particle 
problem and so the results are not surprising. As soon as the masses
are made different, the system becomes nonintegrable and is expected
to have good ergodicity properties \cite{integ}, and correspondingly
the NESS should be very different.  

{\it BBGKY-type equations:}
In principle, a complete solution of the heat conduction problem could be
obtained from the steady-state solution of the master equation for
evolution of the $N$-point distribution equation $\r(\{x_l,u_l\},t)$.
In practice this is difficult and a simpler approach is to work with
the so-called BBGKY hierarchy which deals with reduced distribution
functions \cite{kreu}. 
Let us first make the following
definitions: 
$\r_l(x,u,t)=\la \d{(x-x_l)} \d{(u-u_l)} \ra $,
$\r_{l,l+1}(x_1,u_1;x_2,u_2)=\la \d{(x_1-x_l)}
\d{(u_1-u_l)}\d{(x_2-x_{l+1})}  \d{(u_2-u_{l+1})} \ra$, where $\la A
\ra=\int \r(\{x_l,u_l\},t) A(\{x_l,u_l\}) \prod_l dx_l du_l$. Further
we define $p_{l,l+1}(x,u_1,u_2)$ as the number of 
collisions per unit time occuring at $x$ between the $l$th and
$(l+1)$th  particles with respective velocities $u_1$ and
$u_2$. Clearly
$p_{l,l+1}(x,u_1,u_2)=\r_{l,l+1}(x,u_1;x,u_2) (u_1-u_2) \theta(u_1-u_2)$.
In terms of these the BBGKY-type equations relating one-point functions to
two-point functions are the following:
\bea
&&{\p \r_l(x,u,t)}/{\p t}+u{\p}  \r_l(x,u,t)/{\p x}= \# \\
&&\int p_{l-1,l}(x,u_1,u_2) \d{(u_2'-u)} du_1 du_2-\int
p_{l-1,l}(x,u_1,u) du_1+ \# \\ 
&&\int p_{l,l+1}(x,u_1,u_2) \d{(u_1'-u)} du_1 du_2
-\int p_{l,l+1}(x,u,u_2) du_2.
\label{bbgky}
\eea
These equations hold for $x \neq 0,L$. At the boundaries, the
distribution functions satisfy appropriate boundary conditions.

The physical observables that we will be interested in are the
particle density $n(x,t)$, the energy density $\e(x,t)$ and the
energy current density $j(x,t)$. These can be
expressed in terms of the one-point functions $\r_l(x,u,t)$. Thus we
have:
\bea
n(x,t)&=&\la \sum_l \d(x-x_l) \ra = \sum_l \int \r_l(x,u,t) du \# \\
\e(x,t)&=& \la \sum_l \frac{m_l u_l^2}{2} \d(x-x_l) \ra=\sum_l \frac{m_l}{2}
\int u^2 \r_l(x,u,t) du \# \\
j(x,t)&=& \la \sum_l \frac{m_l u_l^3}{2} \d(x-x_l) \ra=\sum_l \frac{m_l}{2}
\int u^3 \r_l(x,u,t) du 
\eea
The temperature field $T(x)$ is {\it defined} as
$T(x,t)={2 \e(x,t)}/{n(x,t)}$.
Our first result is the following  {\it exact current conservation
equation}: ${\p{\e(x,t)}}/{\p t}+{\p{j(x,t)}}/{\p x}=0$.
This result is easily obtained by multiplying Eq.~(\ref{bbgky}) by $m_l
u^2/2$, integrating over $u$, and summing over all $l$. There is a
pairwise cancellation of all terms on the right-hand side. Similarly
by multiplying Eq.~(\ref{bbgky}) by $m_l u/2$, integrating and summing
gives, in the steady state: 
$\p \e(x)/{\p x}=0 $.
Thus {\it for any choice of masses $\{m_l\}$, the energy density in the
steady state is constant in space}. Physically the constancy of energy
density follows from the constancy of pressure and the linear relation
between the two quantites in an ideal gas. However this does not imply a
constant temperature profile since the temperature also depends on the
number density $n(x)$ which is not constant. 

From the fact that the dynamics [Eq.~(\ref{coll})] is
invariant under a constant scaling of the masses, it follows
that the temperature profile does not change under 
$m_i \to \nu m_i$. Also from the boundary conditions it is easily
shown that $T( \nu T_1, \nu T_2, x )= \nu T(T_1,T_2,x) $.
From now on, we shall consider the dimer case where we consider only
particles of two different masses $m_1$ and $m_2$ placed alternately on
the line. Because of the above scaling properties the only independent
variables are the ratios, $m_2/m_1$ and $T_2/T_1$, and $N$. We will henceforth
consider the case $m_1=1$, $m_2=m$, $T_1=2$ and $T_2=8$.

{\it Numerical results:}
In our numerical simulations we let the system evolve with the appropriate
boundary conditions and compute time-averages of various quantities
in the steady state. The time evolution does not require numerical
solution of differential equations since the exact solution is
essentially known. The system is evolved by 
computing successive collision times and updating velocities using the
collision rules Eq.~(\ref{coll}).
The only errors are those due to 
round-off. We have verified that the simulations reproduce all the exact
known results both for the equal and the unequal mass cases.

In our simulations we vary $\delta=m-1$ and $N$. 
The mass values  $\delta=0.078,0.11,0.22,0.44$ were studied for
particle numbers 
$N=41, 81, 161, 321, 641$ and $1281$. 
The size of the boxes were changed so that in all cases
the average density of particles was fixed at $2$. The
number of particles is chosen to be odd so that at both ends the
particles in contact with the bath have the same mass. The number of
collisions over which the averaging is done was between $10^9-
10^{10}$. In all cases we checked that increasing the time of
averaging by a factor of $10$ did not significantly change the data.

In Fig.~(\ref{del1}), we plot the steady state temperature profiles at
different system sizes for $\delta=0.22$. 
The temperature has a smooth and continuously varying profile. 
There is a jump at the boundaries which decreases as the system size is 
increased and is expected to vanish in the thermodynamic limit when
the current also becomes vanishingly small. 
We find that this is true for 
any non-zero $\delta$. For smaller $\delta$ one needs to go to larger 
system sizes to get the same
temperature profile. Infact for small $\delta$ and large $N$ the
temperature-profile depends on $\delta$ and $N$ only
through the scaling combination $\delta ^2 N$. This is illustrated in
Fig.~(\ref{reduc}) where we plot data corresponding to
five different values of $\delta$, each with a different $N$, chosen
such that $\delta ^2 N$ is the same.  
Note that for $\delta=0$, the temperature profile (which is flat and
given by $T=\sqrt{T_1 T_2}$) and the energy current are both
independent of system size; thus $\delta=0$ is a singular point, while
the steady state in the limit $\delta \to 0,~ 
N \to \infty$ with $ \delta ^2 N $ constant is quite different and nontrivial.

For fixed value of $\delta$, as we increase  $N$ the temperature
profile approaches a 
limiting form. Quite amazingly {\it this limiting form seems to be exactly
one that would be predicted by kinetic 
theory}. We recall that kinetic theory for a one-dimensional gas
predicts Fourier behaviour with $K \sim T^{1/2}$  and hence a
nonlinear temperature profile 
$T_k(x)=[T_1^{3/2} 
(1-x/L) + T_2^{3/2} x/L ]^{2/3}$. This has been plotted in
Fig.~(\ref{del1}). We find that the following scaling form
gives a reasonable collapse of our data:
\bea
T(x,N,\delta)=T_k(x)+\frac{1}{(\delta^2 N)^{\gamma}} g(x).
\label{scaleq}
\eea
The inset in Fig.~(\ref{del1}) shows the collapse of data for
$\delta=0.22$ obtained using the above scaling form with
$\gamma=0.67$.  

We now look at the dependence of $K$ on system size.
In Fig.~(\ref{kvl}) we plot $j$ versus $L$ for 
$\delta=0.22$. 
It is clear from the data that the conductivity which is proportional
to $L j$ has a slow divergence given by $ K \sim L^{\alpha}$
with $\alpha <0.2$. We note
that this is significantly different from the system size dependence
of $K \sim L^{0.4}$ found by Hatano in the same model \cite{hata} and
also in other momentum-conserving systems like FPU and the diatomic Toda
\cite{FPU,hata}.  

Finally we have checked for LTE: to do so we compute the steady state 
expectation $u^{(4)}(x)=\la \sum_l m_l u_l^4 \d(x-x_l) \ra $. If there was LTE, this
quantity would be determined by the local temperature $T(x)$. In
Fig.~(\ref{lte}) we plot $u^{(4)}(x)$, as determined directly by taking
time averages and also the value predicted from the local temperature
$T(x)$. We see that at large system sizes LTE is indeed achieved. We
also find that for smaller values of $\delta$, one needs to go to
larger system sizes to get LTE.  

In summary we have studied heat conduction in the unequal mass problem
which appears to be the simplest nontrivial deterministic system, in
one dimension, for which a very detailed investigation of the NESS can be made.
Our study shows that a meaningful hydrodynamic description of the
steady state is possible even in a situation where (presumably) $K \to
\infty$  in 
the thermodynamic limit.
It is clear that further studies of
this model can throw much light on the difficult problem of transition
from the microscopic to macroscopic description in the context of
nonequilibrium phenomena.  

I thank D. Dhar and M. Rao for many valuable suggestions. I also thank 
J. Das, C. Dasgupta, R. Pandit, S. Ramaswamy  and B. S. Shastry for
discussions. 

$^{\dagger}$ Also at the Poornaprajna Institute, Bangalore. dabhi@rri.res.in

\vspace{2cm}
\vbox{
\epsfxsize=8.0cm
\epsfysize=6.0cm
\epsffile{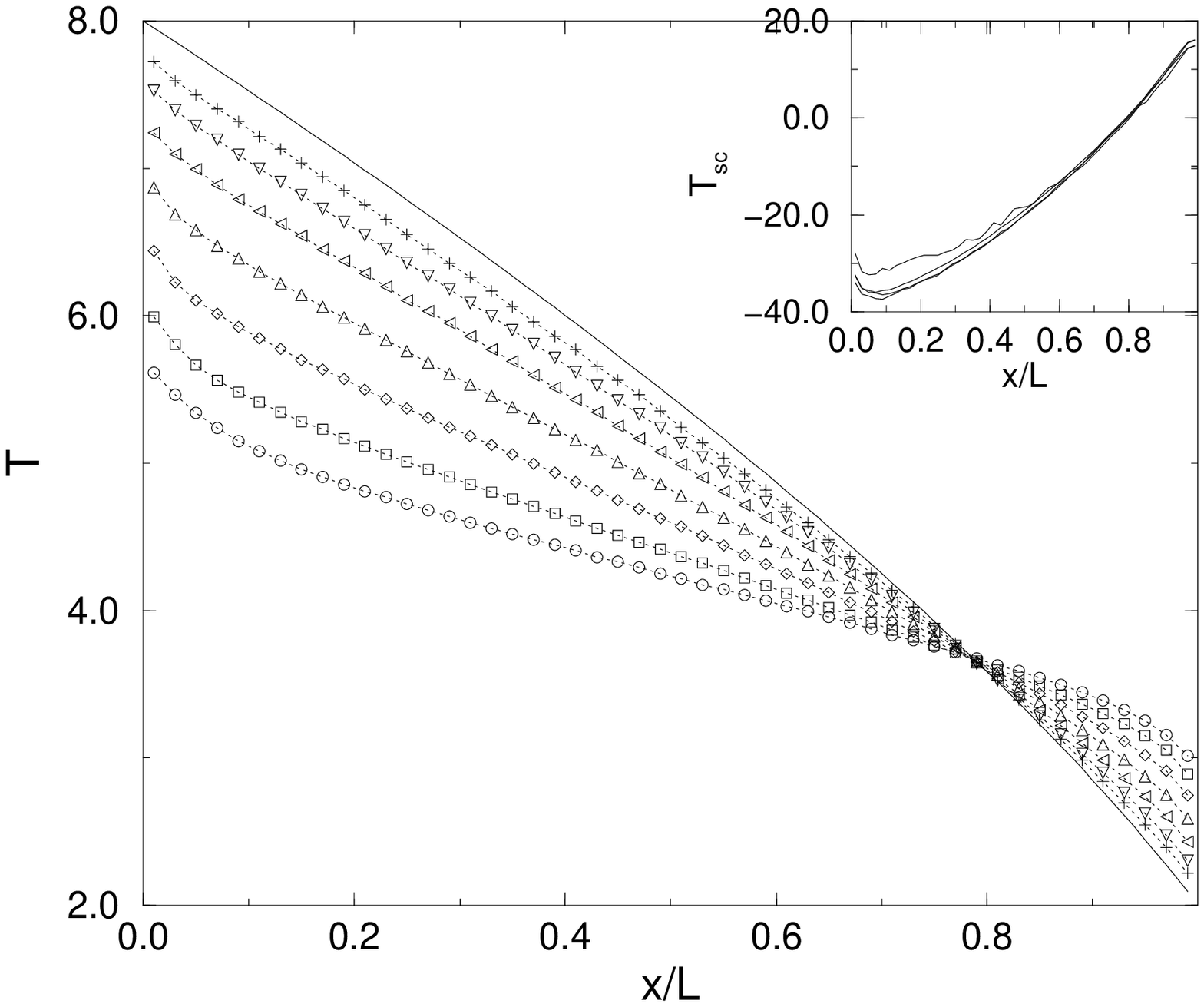}
\begin{figure}
\caption{\label{del1} 
Temperature profiles for $\delta=0.22$ are
plotted for system sizes $N=21 ({\rm{indicated~ by~ o}})
,41,81,161,321,641$ and $1281~ (+)$. The solid line is the prediction
of kinetic theory.  
In the inset we have plotted $T_{sc}=N^{0.67} (T(x)-T_k(x)$ [see
Eq.~(\ref{scaleq})] with the data for $N=161,321,641$ and $1281$. 
}  
\end{figure}}
\vbox{
\epsfxsize=8.0cm
\epsfysize=6.0cm
\epsffile{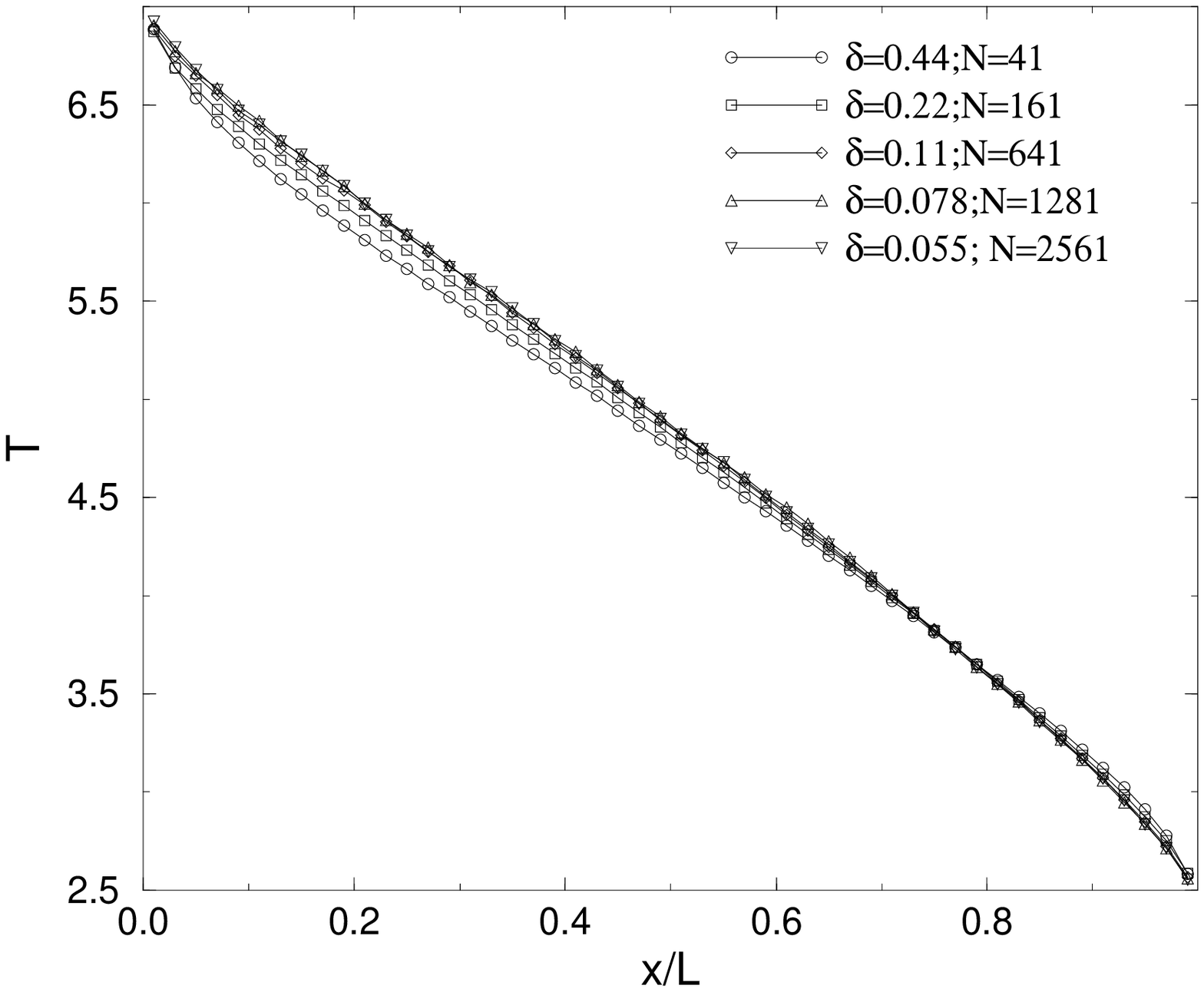}
\begin{figure}
\caption{\label{reduc} Temperature profiles obtained with five different
sets of values for $\delta$ and $N$ with $\delta^2 N$ constant.   
}  
\end{figure}}
\vbox{
\epsfxsize=8.0cm
\epsfysize=6.0cm
\epsffile{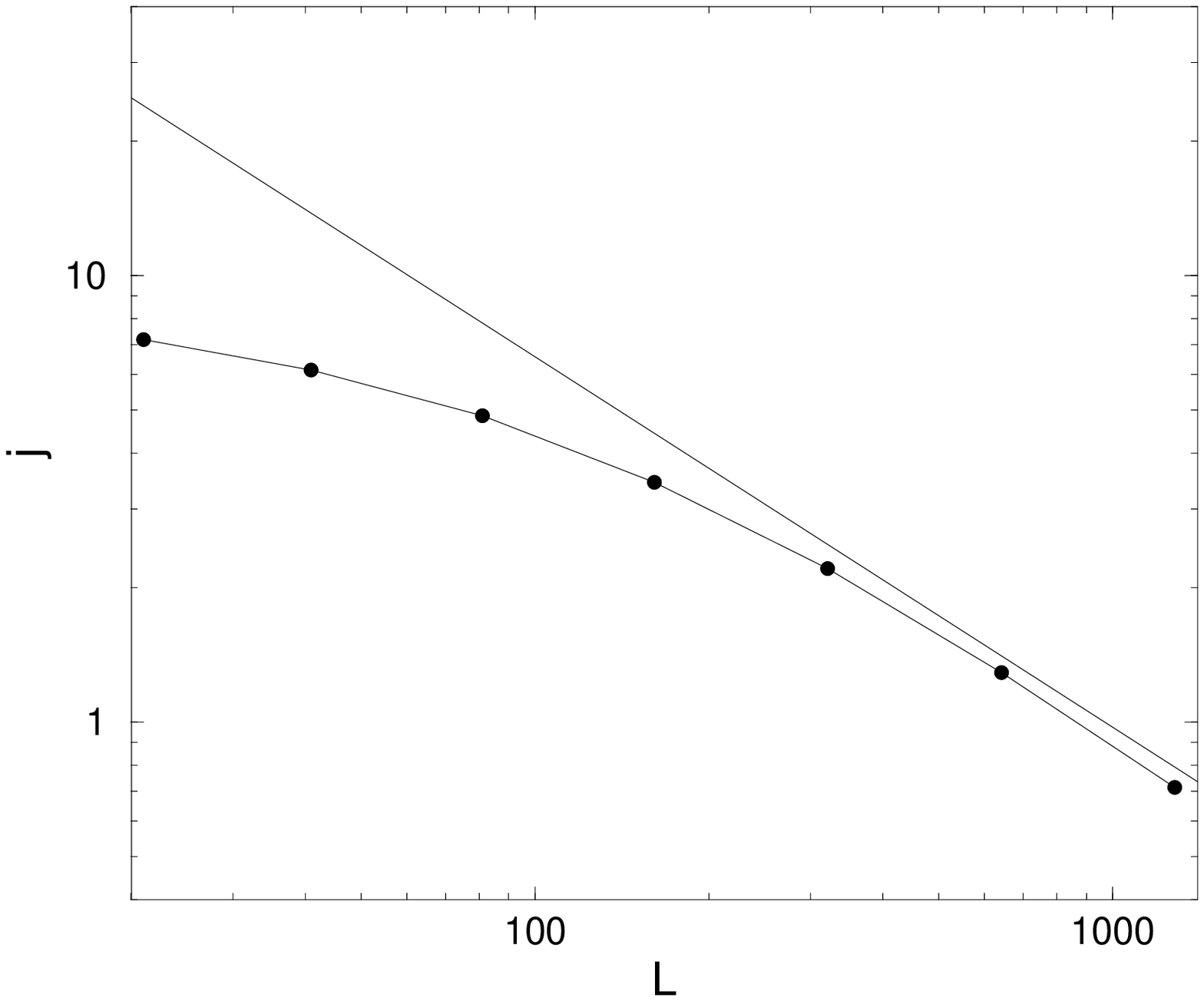}
\begin{figure}
\caption{\label{kvl} Plot of $j$ versus $L$ for $\delta=0.22$ The
straight line shown corresponds to the decay $j\sim 1/L^{0.83}$.   
}  
\end{figure}}
\vbox{
\epsfxsize=8.0cm
\epsfysize=6.0cm
\epsffile{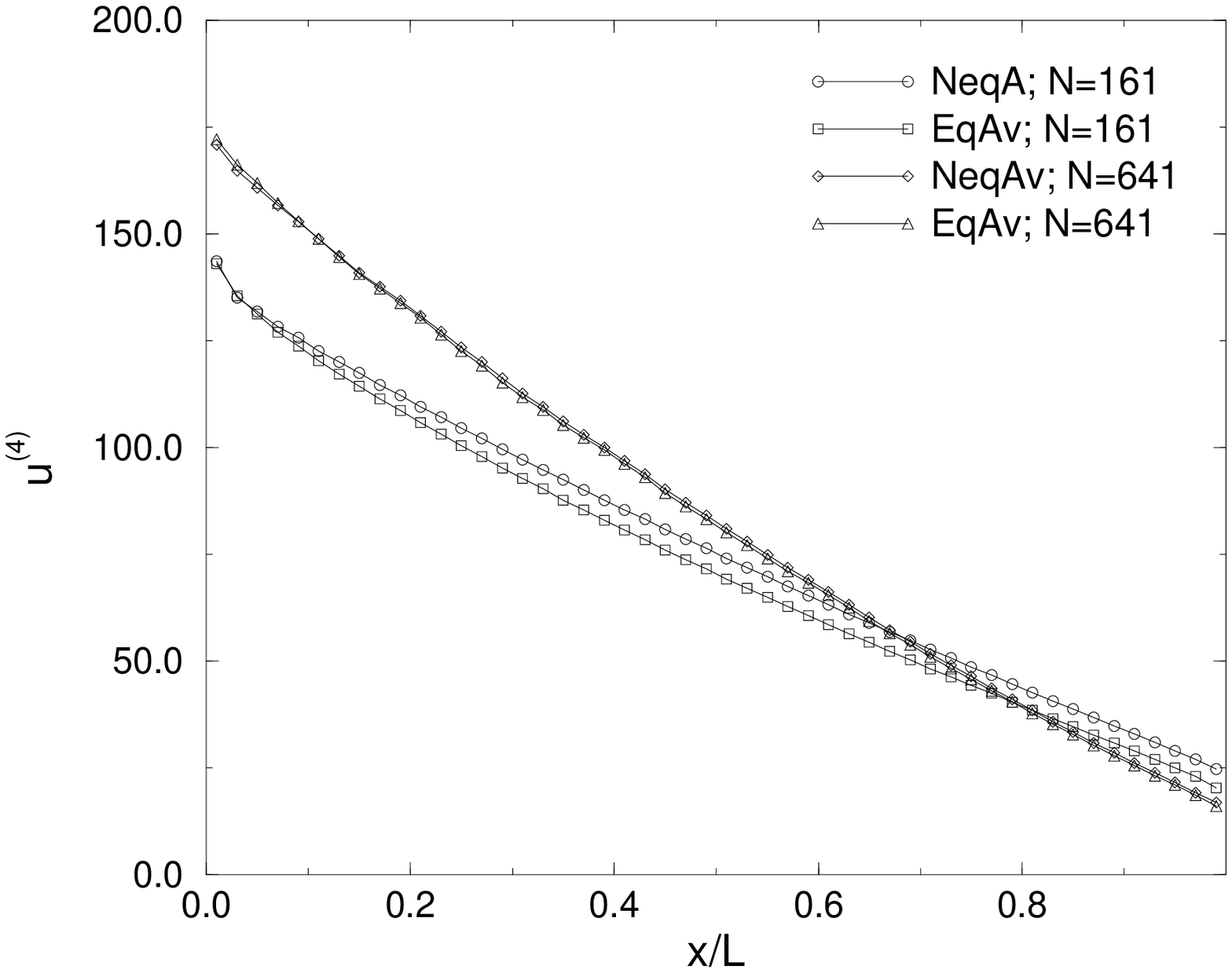}
\begin{figure}
\caption{\label{lte} Plot of $u^{(4)}$ as determined from a direct time
averaging (NeqAv) and from the local temperature (EqAv) for two system
sizes. For the bigger system, the curves almost coincide. 
}  
\end{figure}}
 

\begin{references}
\bibitem{fnote} It is quite interesting to note
that the $T^{1/2}$ dependence of the conductivity, for arbitrary
masses, can also be obtained from the Kubo formula by
simply scaling out the temperature.
\end{references}
\end{document}